\documentclass[8pt]{article}
\usepackage{graphicx,psfrag}
\usepackage{hyperref}
\usepackage{amssymb}
\usepackage{amsmath}
\usepackage{amscd}
\usepackage{amsthm}
\usepackage[font=small,labelfont=bf,justification=centerlast]{caption}
\usepackage{subcaption}

\usepackage{enumerate}

\usepackage[left=2cm,top=2cm,right=2cm,bottom=2cm]{geometry}

\begin{document}
Gravitation and Cosmology (2013) \textbf{00}: 00-00\\

\begin {center}\Large
\textbf{Anisotropic model of dark energy dominated universe\\ with hybrid expansion law
}\\
\end {center}

\begin{center}
\textbf{Suresh Kumar \footnote{E-Mail: sukuyd@gmail.com, suresh.kumar@pilani.bits-pilani.ac.in}}\end {center}
\begin{center}
Department of Mathematics, BITS Pilani, 
Pilani Campus, Rajasthan-333031, India. 
\end {center}

\noindent\textbf{Abstarct:}
The paper deals with the study of the dynamics of Universe within the framework of a spatially homogeneous Bianchi-V space-time filled with a perfect fluid composed of non-interacting matter and dynamical dark energy components. We determine the Bianchi-V space-time by considering hybrid expansion law (HEL) for the average scale factor that yields power-law and exponential-law cosmologies in its special cases. In the HEL cosmology, the 
Universe exhibits transition from deceleration to acceleration. We find that the HEL Universe within the framework of Bianchi-V space-time is anisotropic at the early stages of evolution and becomes isotropic at late times. The dynamical dark energy in the HEL Bianchi-V Universe does not show departure from the usual cosmological constant at later times.


\section{Introduction}
The phenomenon of accelerated expansion of the Universe came into picture in the last decade, and  by now it has been confirmed by different
data sets of complimentary nature such as type Ia supernovae (SN
Ia), baryon oscillations, galaxy clustering, cosmic microwave
background (CMB) and weak lensing \cite{1}. But this has posed a challenging problem- what is driving the accelerated expansion of the Universe. After estimating various energy components of Universe, the cause of accelerated expansion of the Universe has been attributed to some
exotic energy stuff dubbed as the dark energy (DE). The most
recent Planck results indicate that DE is major ingradient in the energy budget of the Universe and accounts for nearly 70\% of the
total mass energy of the Universe \cite{2}. It is believed that DE possesses negative pressure and tends to
increase the rate of expansion of the Universe \cite{3}.  Theoretically, the so called cosmological constant ($\Lambda$), once abandoned by Einstein, is put forward as a simplest candidate of DE, characterized by the equation of state (EoS) $p^{(DE)}=-\rho^{(DE)}$.
However, the cosmological constant suffers from the theoretical problems such as the ``fine-tuning" and ``cosmic coincidence" puzzles \cite{4}. This prompted theorists to study cosmological models with dynamical DE characterized by an effective EoS $w^{(DE)}=p^{(DE)}/\rho^{(DE)}\neq-1$ \cite{5, 6}.  

On larger scales our Universe is isotropic and homogeneous and is well described observationally by the $\Lambda$CDM model within the framework of Friedmann-Robertson-Walker (FRW) geometry as demonstrated in the recent Planck data studies \cite{2}. However, $\Lambda$CDM cosmology shows a poor fit to the CMB temperature power spectrum at low multipoles \cite{2}.
This indicates that the isotropy and homogeneity were not the essential features of the early Universe. 
Therefore, the recent Planck data results \cite{2} motivate us to study models with anisotropic geometry for a better understanding of the evolution of the Universe.
In this regard, Bianchi type V space-time is of fundamental
importance since it provides the requisite framework. A spatially
homogeneous and anisotropic Bianchi type-V model is considered as
the simplest generalization of the Bianchi type-I and flat FRW models. It is described
by the line element (in units $c=1$)
\begin{equation}\label{1}
ds^{2} =-dt^{2} +A^{2}dx^{2} +e^{2m x}(B^{2}dy^{2}
+C^{2}dz^{2}),
\end{equation}
where $A$ , $B$ and $C$ are the metric functions or directional
scale factors of cosmic time $t$, and $m$ is a constant. When $m=0$, the space-time reduces to the Bianchi type-I model. Further if all the three scale factors are equal, that is, $A=B=C$ with $m=0$, the space-time reduces to the standard flat FRW model. The average scale factor of the Bianchi-V Universe is defined as $a=(ABC)^{1/3}$.  

In Ref.\cite{Kumaranil11}, we studied power-law ($a(t)\propto t^\alpha$) and exponential-law ($a(t)\propto e^{\beta t} $) cosmologies within the framework of Bianchi-V models with non-interacting matter fluid and DE components. Thereafter in another study \cite{Kumar12}, we investigated various features of power-law cosmology by constraining it with a host of observational data and found that such a cosmology is not a complete package for cosmological purposes. In fact, power-law and exponential-law cosmologies can be used only to describe epoch based evolution of the Universe because of the constancy of deceleration parameter. For instance, these cosmologies do not exhibit the transition of the Universe from deceleration to acceleration.  In a recent paper \cite{HEL13}, we considered the following anastz for the scale factor of the Universe:
\begin{equation}\label{2}
a(t)=a_1 t^{\alpha}e^{\beta t} ,
\end{equation}
where $a_1>0$, $\alpha\geq 0$ and $\beta \geq 0$ are constants. We referred this generalized form of scale factor to as the Hybrid Expansion Law (HEL) being the mixture of power-law and exponential-law cosmologies. One may immediately observe that the HEL leads to the power-law cosmology for $\beta=0$ and to the exponential-law cosmology for $\alpha=0$. In other words, the power-law and exponential-law cosmologies are the special cases of the HEL cosmology. Therefore, the case $\alpha>0$ and $\beta>0$ leads to a new cosmology arising from the HEL. In the Ref. \cite{HEL13}, we determined the scalar fields (quintessence, phantom, tachyon) with potentials responsible for the evolution of the Universe under the HEL within the framework of FRW spacetime in general relativity. We confronted the HEL cosmology with the latest observational data from H(z) and SN Ia compilations, and then studied the kinematics and dynamics of the HEL Universe in detail. 

In this paper, our intention is to study anisotropic evolution of the Universe dominated by dynamic dark energy and driven by the HEL. Therefore, we choose the fabric of Bianchi-V space-time \eqref{1} filled with non-interacting matter (species of matter of gravitating nature such as baryonic, cold dark matter etc.) and DE components. Further, we allow the EoS parameter $w^{(DE)}$ of DE to vary with time so that we can study dynamical nature of DE and examine its departure, if any, from the usual cosmological constant. In Section II, we present solution to the field equations related to Bianchi-V spacetime assuming the constancy of the EoS parameter of non-interacting matter component and taking into account the HEL. In Section III, we discuss kinematics and physical behavior of the derived model. Concluding remarks are written in Section IV.

\section{Field equations and their solution}
The Einstein's field equations in case of a mixture of perfect fluid
and DE components, in the units $8\pi G=c=1$,  read as
\begin{equation}\label{3}
R_{ij}-\frac{1}{2} g_{ij}R =- T_{ij},
\end{equation}
where $T_{ij}=T^{(M)}_{ij}+T^{(DE)}_{ij}$ is the overall energy
momentum tensor with $T^{(M)}_{ij}$ and $T^{(DE)}_{ij}$ as the
energy momentum tensors of matter and DE, respectively.
These are given by
\begin{equation}\label{4}
T^{(m)~i}_{~j}=diag[-\rho^{(M)},p^{(M)}\;,p^{(M)},p^{(M)}]
                   =diag[-1,w^{(M)},w^{(M)},w^{(M)}]\rho^{(M)}
\end{equation}
and
\begin{equation}\label{5}
T^{(de)~i}_{~j}=diag[-\rho^{(DE)},p^{(DE)},p^{(DE)},p^{(DE)}]
                   =diag[-1,w^{(DE)},w^{(DE)},w^{(DE)}]\rho^{(DE)}
\end{equation}
where $\rho^{(M)}$ and $p^{(M)}$ are, respectively the energy
density and pressure of the matter while $w^{(M)}=p^{(M)}/\rho^{(M)}$ is its EoS
parameter. Similarly,  $\rho^{(DE)}$ and $p^{(DE)}$ are,
respectively the energy density and pressure of the DE component
while $w^{(DE)}=p^{(DE)}/\rho^{(DE)}$ is the corresponding EoS
parameter.

In a comoving coordinate system, the field equations (\ref{3}),
for the Bianchi-V space-time (\ref{1}), in case of (\ref{4}) and
(\ref{5}), read as
\begin{equation}\label{6}
\frac{\ddot{B}}{B} +\frac{\ddot{C}}{C}
+\frac{\dot{B}\dot{C}}{BC}-\frac{m^{2}}{A^{2}}=-w^{(M)}\rho^{(M)}-w^{(DE)}\rho^{(DE)},
\end{equation}
\begin{equation}\label{7}
\frac{\ddot{C}}{C} +\frac{\ddot{A}}{A}
+\frac{\dot{C}\dot{A}}{CA}-\frac{m^{2}}{A^{2}}
=-w^{(M)}\rho^{(M)}-w^{(DE)}\rho^{(DE)},
\end{equation}
\begin{equation}\label{8}
\frac{\ddot{A}}{A} +\frac{\ddot{B}}{B}
+\frac{\dot{A}\dot{B}}{AB}-\frac{m^{2}}{A^{2}}
=-w^{(M)}\rho^{(M)}-w^{(DE)}\rho^{(DE)},
\end{equation}
\begin{equation}\label{9}
\frac{\dot{A}\dot{B}}{AB} +\frac{\dot{B}\dot{C}}{BC}
+\frac{\dot{C}\dot{A}}{CA}-\frac{3m^{2}}{A^{2}}
 =\rho^{(M)}+\rho^{(DE)}.
\end{equation}
\begin{equation}\label{10}
2\frac{\dot{A}}{A}-\frac{\dot{B}}{B}
-\frac{\dot{C}}{C}=0
\end{equation}
The energy conservation equation $T^{ ij}_{~ ;j} =0$ yields
\begin{equation}\label{11}
\dot\rho^{(M)}+3(1+w^{(M)})\rho^{(M)}H+
\dot\rho^{(DE)}+3(1+w^{(DE)})\rho^{(DE)}H=0,
\end{equation}
where $H=\frac{1}{3}\left(H_x+H_y+H_z\right)$ is the average Hubble parameter with $H_x=\frac{\dot{A}}{A}$, $H_y=\frac{\dot{B}}{B}$ and $H_z=\frac{\dot{C}}{C}$ defined as the directional Hubble parameters.

The energy momentum tensors of non-interacting matter and DE fluids can be conserved separately. The energy conservation equations of matter and DE, respectively, read as

\begin{equation}\label{12}
\dot\rho^{(M)}+3(1+w^{(M)})\rho^{(M)}H=0,
\end{equation}

\begin{equation}\label{13}
\dot\rho^{(DE)}+3(1+w^{(DE)})\rho^{(DE)}H=0,
\end{equation}

Next, the EoS parameter $w^{(M)}$ of the matter fluid is assumed to be a constant while $w^{(DE)}$ is allowed to vary with time in order to study dynamical nature of DE. In view of constancy of $w^{(M)}$, integration of \eqref{12} leads to
\begin{equation}\label{14}
\rho^{(M)}=c_{0}a^{-3(1+w^{(M)})},
\end{equation}
where $c_{0}$ is a positive constant of integration.

From equations (\ref{6})-(\ref{10}), the metric functions can be explicitly written in terms of the the average scale factor $a$ as (see \cite{Kumaranil11} for details)
\begin{equation}\label{15}
A(t)= a ,
\end{equation}
\begin{equation}\label{16}
B(t)=h a \exp \left(l \int a^{-3} dt \right),
\end{equation}
\begin{equation}\label{17}
C(t)=h^{-1} a \exp \left(-l \int a^{-3} dt \right),
\end{equation}
where
$h$ and $l$ are constants.

Using the HEL (\ref{2}), we find the metric functions explicitly in terms of $t$ as follows 
\begin{equation}\label{18}
A(t)= a_1t^{\alpha}e^{\beta t} ,
\end{equation}
\begin{equation}\label{19}
B(t)=h a_1t^{\alpha}e^{\beta t} \exp \left(-l a_1^{-3}(3\beta)^{3\alpha-1} \gamma[1-3\alpha, 3\beta t] \right),
\end{equation}
\begin{equation}\label{20}
C(t)=h^{-1} a_1t^{\alpha}e^{\beta t} \exp \left(l a_1^{-3}(3\beta)^{3\alpha-1} \gamma[1-3\alpha, 3\beta t] \right),
\end{equation}
where $\gamma$ denotes the lower incomplete gamma function. We find the condition $\alpha\leq 1/3$ for the metric functions $B(t)$ and $C(t)$ to be realistic.

The energy density of matter component in terms of time is
\begin{equation}
\rho^{(M)} =c_{0}a_1^{-3(1+w^{(M)})}t^{-3\alpha(1+w^{(M)})}e^{-3\beta(1+w^{(M)}) t}.
\end{equation}

The energy density and EoS parameter of DE respectively read as
\begin{equation}
\rho^{(DE)} = 3(\alpha t^{-1}+\beta)^{2}-l^2 a_1^{-6}t^{-6\alpha}e^{-6\beta t}-3m^2 a_1^{-2}t^{-2\alpha}e^{-2\beta t}-c_{0}a_1^{-3(1+w^{(M)})}t^{-3\alpha(1+w^{(M)})}e^{-3\beta(1+w^{(M)}) t},
\end{equation}
\begin{equation}
w^{(DE)} = -1-\frac{2}{\rho^{(DE)}}\left[-\alpha t^{-2}+l^2 a_1^{-6}t^{-6\alpha}e^{-6\beta t}+m^2 a_1^{-2}t^{-2\alpha}e^{-2\beta t}+c_{1}t^{-3\alpha(1+w^{(M)})}e^{-3\beta(1+w^{(M)}) t}\right],
\end{equation}
where $c_1=(c_0/2)(1+w^{(M)}) a_1^{-3(1+w^{(M)})}$.

Thus, we have obtained explicit expressions in terms of $t$ for all the variable ingredients  $A(t)$, $B(t)$, $C(t)$, $\rho^{(M)}$, $w^{(DE)}$, $\rho^{(DE)}$ of the field equations (\ref{6})-(\ref{10}). 
\section{Kinematics and physical behavior of the model}
The deceleration parameter of the model is
\begin{equation}\label{}
q=-\dfrac{a\ddot{a}}{\dot{a}^2}=\dfrac{\alpha }{(\beta t+\alpha )^{2}}-1.
\end{equation}

We observe that the HEL Universe evolves with variable deceleration parameter and the transition from deceleration to acceleration takes place at 
\begin{equation}\label{}
t=\dfrac{\sqrt{\alpha}-\alpha}{\beta},
\end{equation}
which restricts  $\alpha$ in the range $0<\alpha<1$.

The directional Hubble parameters $H_x$, $H_y$, $H_z$ and the average Hubble parameter $H$ are given by
\begin{equation}\label{}
H_x=\alpha t^{-1}+\beta,\;H_y=\alpha t^{-1}+\beta +l a_1^{-3}t^{-3\alpha}e^{-3\beta t},\;H_z=\alpha t^{-1}+\beta-a_1^{-3}t^{-3\alpha}e^{-3\beta t},\;H=\alpha t^{-1}+\beta .
\end{equation}

The anisotropy parameter ($\bar{A}$) and shear scalar $(\sigma)$ of
the model read as
\begin{equation}\label{}
\bar{A}=\frac{1}{9H^{2}}\left[(H_x-H_y)^{2}+(H_y-H_z)^{2}+(H_z-H_x)^{2}
\right]=\dfrac{l^2}{9}(\alpha t^{-1}+\beta)^{-2} a_1^{-6}t^{-6\alpha}e^{-6\beta t},
\end{equation}
\begin{equation}\label{}
\sigma^{2}=\frac{3}{2}\bar{A}H^{2}=\dfrac{l^2}{6}a_1^{-6}t^{-6\alpha}e^{-6\beta t}.
\end{equation}

At this stage, it is worthwhile to note that in the HEL cosmology, the observational setting $a(t)=1/(1+z)$, $z$ being the redshift, leads to the following relation between time and redshift:
\begin{equation}\label{25}
t=\left(\frac{\alpha}{\beta}\right)\text{W}\left[\frac{\beta}{\alpha} \left(\frac{1}{a_1 (1+z)}\right)^{\frac{1}{\alpha}}\right],\end{equation}
 where $W$ denotes the Lambert W function, also known as the omega function or product logarithm.
Using relation \eqref{25}, one may express parameters of the derived model in terms of redshift. Such a relation is useful for testing the model with observational data. Also, one has the liberty to test the behavior of the parameters with respect to cosmic time or redshift. For an illustration, in Fig. 1(a) we show the variation of $q(z)$ versus $z$ and in Fig.1(b) we show the variation of $w^{(DE)}$ versus $z$ for selected values of the constants given in the caption of the figure.
\begin{figure*}[htb!]
        \centering
        \begin{subfigure}[b]{0.45\textwidth}
                \centering
                 \psfrag{q(z)}[b][b]{$q(z)$}
				\psfrag{z}[b][b]{$z$}
                \includegraphics[width=\textwidth]{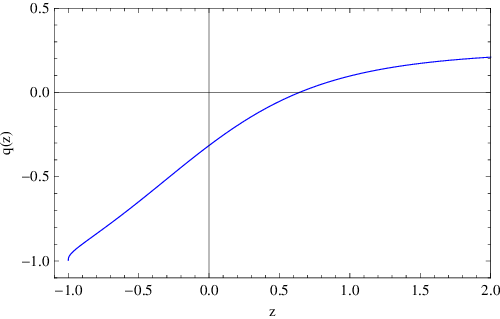}
                \caption{ }
                \label{fig1a}
        \end{subfigure}
        \hspace{0.02\linewidth}
        \begin{subfigure}[b]{0.45\textwidth}
                \centering
                \psfrag{wde}[b][b]{$w^{(DE)}$}
				\psfrag{z}[b][b]{$z$}
                \includegraphics[width=\textwidth]{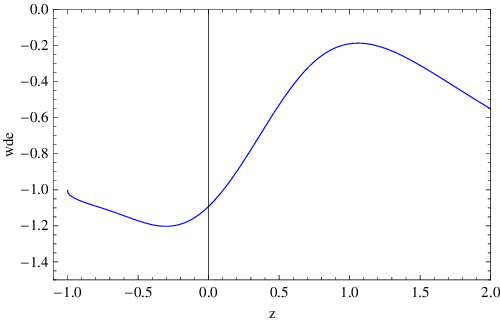}
				\caption{ }                
                \label{fig1b}
        \end{subfigure}
\caption{\footnotesize{ \textbf{(a)} Variation of $q(z)$ versus $z$ for $\alpha=0.29$, $\beta=0.48$ and $a_1=0.95$. The transition of the Uinverse from deceleration to acceleration takes place at $z=0.51$ and the present value of DP is $q(0)=-0.31$.\textbf{(b)} Variation of $w^{(DE)}$ versus $z$ for $\alpha=0.29$, $\beta=0.48$, $a_1=0.95$, $l=0.88$, $m=0.5$, $c_0=0.12$. The present value of EoS parameter of DE is $w^{(DE)}(0)=-1.09$. }}
\end{figure*}

\begin{table}[h]\centering\small
\caption{Asymptotic behavior of the derived model parameters.} 
\begin{tabular}{|c|c|c|}
\hline Parameters & $t\rightarrow 0$ ($z\rightarrow \infty$)& $t\rightarrow \infty$ ($z\rightarrow -1$) \\ \hline
$A$, $B$, $C$, $a$ & $0$     &$\infty$\\[8pt]
$\rho^{(M)}$ & $\infty$     &$0$\\[8pt]
$\rho^{(DE)}$ & Indeterminate     &$3\beta^2$\\[8pt]
$w^{(DE)}$ & Indeterminate     &$-1$\\[8pt]
$q$ & $\frac{1}{\alpha}-1$     &$-1$\\[8pt]
$H$, $H_x$, $H_y$, $H_z$ & $\infty$     &$\beta$\\[8pt]
$\bar{A}$, $\sigma^2$& $\infty$     &$0$\\[8pt]
\hline
\end{tabular}
\label{table:HELobs}
\end{table}

In Table I, we give the asymptotic behavior of the derived model parameters. We see that the spatial volume $a^3$ and the directional scale factors $A$, $B$, $C$ vanish while the other parameters such as $\rho^{(M)}$, $H$, $H_x$, $H_y$, $H_z$, $\bar{A}$ and $\sigma^{2}$ diverge as $t\rightarrow 0$. Thus, the model exhibits singularity  at $t=0$ \cite{Bronnikov2004}. Further, we observe that the Universe starts evolving with different expansion rates $H_x$, $H_y$, $H_z$ along x, y and z directions. The Universe has high anisotropy and shear in the beginning. But with sufficient growth of the scale factor with time, we have
\[H_x\sim H_y \sim H_z\sim \beta,\;\;\;\;\bar{A}\sim 0,\;\;\;\;\sigma\sim 0.\] This shows that anisotropy and shear become negligible, and the Universe tends to isotropy at late times in the derived model \cite{Bronnikov2004}. This is consistent with the astronomical observations which advocate isotropic Universe on larger scales. 

Next, for sufficiently large times, we find
\[H \sim \beta,\;\;\;\;q\sim -1,\;\;\;\;\rho^{(M)}\sim 0,\;\;\;\;\rho^{(DE)}\sim 3\beta^2,\;\;\;\;w^{(DE)}\sim -1.\]

Thus, the HEL Universe asymptotically achieves the de Sitter phase and hence expands forever with the dominance of DE in the form of cosmological constant.

\section{Concluding remarks}
In this paper, we have studied dynamics of the Universe within the framework of a spatially homogeneous Bianchi-V space-time filled with a perfect fluid composed of non-interacting matter and dynamical dark energy components. We have determined the Bianchi-V space-time by considering HEL for the average scale factor that yields power-law and exponential-law cosmologies in its special cases. We find that the HEL Universe exhibits transition from deceleration to acceleration which is an essential feature of dynamic evolution of the Universe. The Bianchi-V HEL Universe begins with high anisotropy but becomes isotropic at the later stages of the evolution. We also find that the dark energy in the HEL Bianchi-V Universe is dynamical in nature. However, it does not show any departure from the usual cosmological constant at later times. Thus, in the derived model, the Universe evolves to the de Sitter phase as it does in the standard $\Lambda$CDM model. Finally, studies of the theoretical anisotropic models such as the one carried out in this paper may be fruitful while dealing with the issues of CMB anisotropy, structure formation in the early Universe etc.
\subsection*{Acknowledgments}
The author acknowledges the warm hospitality and research facilities provided by the Inter-University Centre for Astronomy and Astrophysics (IUCAA), India where a part of this work was carried out. Sincere thanks are also in place to the anonymous referee for constructive comments on the manuscript.

\end{document}